\documentclass[aps,prl,twocolumn,groupedaddress]{revtex4}
\usepackage{graphicx,amsmath}
\begin{document}

\title{Understanding spin glass transition as a dynamic phenomenon}
\author{Kostya Trachenko$^{1,2}$}
\address{$^1$ SEPnet and School of Physics, Queen Mary University of London, Mile End Road, London, E1 4NS, UK}
\address{$^2$ Department of Earth Sciences, University of Cambridge, Cambridge CB2 3EQ, UK}
\begin{abstract}
Existing theories explain spin glass transition in terms of a phase transition and order parameters, and assume the existence of a distinct spin glass phase. In addition to problems related to clarifying the nature of this phase, the common challenge is to explain profound dynamic effects. Here, we propose that the main experimental results of spin glass transition can be understood in an entirely dynamic picture, without a reference to a distinct spin glass phase, phase transition and order parameters. In this theory, the susceptibility cusp at the glass transition temperature is due to the dynamic crossover between the high-temperature relaxational and low-temperature spin wave regime. The crossover takes place when $t=\tau$, where $t$ is observation time and $\tau$ is relaxation time. Time-dependent effects, inconsistent with the phase transition approach, and the logarithmic increase of $T_g$ with field frequency in particular, originate as the immediate consequence of the proposed picture. In our discussion, we explore similarities between the spin and structural glass transitions.
\end{abstract}


\maketitle

\section{Introduction}

Since its discovery, spin glass has been considered as a third distinct type of low-temperature magnetic arrangement in solids, in addition to ferromagnetic and anti-ferromagnetic \cite{mydosh}. In spin glass, spins are disordered, prompting the term ``glass'', and give zero net magnetic moment. Great theoretical effort went into understanding the nature of the spin glass transition. The collection of new ideas and theories has formed a new large research field in condensed matter physics, with connections proposed to other disciplines such as economics and biology \cite{mydosh,s2,s3}. A couple of decades ago, it was remarked that ``as fame and topicality diminish, the problem remains an interesting, yet pass\'{e}, research topic'' \cite{mydosh}. One of the reasons for continued interest in this area is the presence of fundamental theoretical problems that are not resolved. There is also current interest in new systems such as magnetic nanoparticles, ferrofluids, relaxor ferroelectrics and so on, where spin glass behavior is observed.

The main signature of the spin glass transition is the susceptibility cusp, which is considered to mark the transition from the paramagnetic to the spin glass phase \cite{mydosh}. The temperature at which the cusp develops is called spin glass transition temperature, $T_g$ (or $T_f$). The susceptibility cusp has stimulated approaches to spin glass transition based on the existence of a phase transition at $T_g$, between the high-temperature paramagnetic phase and a novel low-temperature ``spin glass'' phase. This was primarily inspired by the phase transition theory, centered at the changes of system properties at the phase transition temperature \cite{landau,ma}.

The traditional phase transition theory requires the presence of two distinct phases on both sides of the phase transition \cite{landau,ma}. On the other hand, as detailed studies revealed, there is no obvious distinct second phase in the spin glass transition: the structure of the low-temperature spin-glass phase is not different from the high-temperature paramagnetic phase, and is equally disordered. The absence of a second phase presents a problem with viewing spin glass transition as a phase transition, resulting in persisting difficulties and controversies \cite{mydosh}. To circumvent the problem, popular theories introduced and discussed novel non-conventional order parameters \cite{mydosh,s2,s3} while retaining the view that a phase transition of some sort takes place. An advantage of this approach is that well-established methods of statistical mechanics can be used to calculate system equilibrium properties and their derivatives such as susceptibility. Such was the Edwards-Anderson (EA) theory \cite{edwards} which introduced the new order parameter for spin glass: $q=\langle s_i^{(1)}s_i^{(2)}\rangle$, where superscripts $1$ and $2$ denote two different moments of time for spin $i$.

As introduced, $q$ serves to quantify the difference between the mobile spins in the paramagnetic phase and the frozen spins in the glass phase at $T_g$. Notably, $q$ describes the change in dynamic behaviour but not the difference between two distinct equilibrium phases as in the conventional phase transition theory. The important next step in the EA theory is the substitution of the time average by the ensemble average, by making an assumption that the system is in equilibrium and is ergodic. Subsequently, statistical mechanics is used to calculate system's equilibrium thermodynamic functions.

The EA theory was followed by numerous theoretical studies that introduced new ideas, concepts and novel mathematical frameworks aimed at clarifying and understanding the unconventional nature of the spin-glass phase transition, the low-temperature phase and the EA order parameter \cite{mydosh,s2,s3}. Importantly, similar to the EA theory, current approaches are based on the existence of a second distinct spin glass phase and a phase transition of some sort, and discuss the associated order parameters. Consequently, the predominant current view is that spin glass transition is a thermodynamic phase transition, even though unconventional \cite{mydosh,s2,s3,binder,binder1}.

Several outstanding problems remain with the phase transition approach. An essential problem rests with identifying the nature of the spin glass phase. This begs the important question of whether experimental data can be understood without making the assumption that a distinct spin glass phase exists. Another important issue is the conflict between theory and experiments \cite{mydosh}. Experiments widely show strong dynamic effects. $T_g$ significantly increases with field frequency, or as observation time decreases \cite{mydosh,v6,v2,v5,v1,v9,v7,v3,v4,v8,male,bhow1,stringer}. This is inconsistent with a transition between two equilibrium phases \cite{mydosh}. We note at this point that the same effect is seen in the liquid-glass (structural) transition \cite{loga1,loga2}.

In addition to the susceptibility cusp, other properties of spin glass systems have been studied, but have not lead to a conclusive evidence for a phase transition \cite{mydosh,s2,binder}. For example, magnetic heat capacity shows no anomaly at $T_g$, unlike in a phase transition \cite{mydosh,s2}. On the other hand, the behavior of non-linear susceptibility, $\chi_{nl}$, above $T_g$ is interpreted to be consistent with critical scaling and is taken as evidence for the static phase transition at $T_g$ \cite{binder}. This conclusion is shared by most authors \cite{s2}, although others are more cautious \cite{mydosh}. It is noted that scaling exists away from $T_g$ only, but fails close to $T_g$ where $\chi_{nl}$ flattens off, inconsistent with a phase transition picture \cite{mydosh,binder}. There are other problems with analyzing this and other similar scaling data: first, the choice of $T_g$ is arbitrary because $T_g$ is not fixed but depends on observation time, introducing arbitrariness in the scaling analysis. This applies not only to the interpretation of experiments, but also to extensive simulations of spin glasses \cite{binder}. In particular, as larger times are accessed in the simulations, $T_g$ decreases as expected from the experiments, altering critical exponents, lower critical dimension and so on \cite{binder1}, and thus suggesting that no firm values of these quantities can be asserted. Second, a phase transition is accompanied by the scaling behavior, but the opposite does not generally need to be the case. Consequently, the scaling behavior is seen in spin glass simulations where the phase transition is known to be absent \cite{s2,binder}, weakening the evidence for a phase transition from the scaling data. Finally, the system is not in equilibrium at and below $T_g$, so it is not known that the scaling data are related to a new phase \cite{mydosh,s2,binder}. We note that slow non-equilibrium relaxation effects could precede a true phase transition as well (eg paramagnetic to ferromagnetic), however in this case the presence of the distinct low-temperature phase is known experimentally.

The non-equilibrium state of the spin glass system generally complicates the approach based on a phase transition. In both experiments and simulations, relaxation time $\tau$ continuously increases on lowering the temperature \cite{mydosh,s2,binder}, as is the case in the structural glass transition. At a certain temperature, $\tau$ exceeds observation time, at which point the system falls out of equilibrium. This presents a problem for understanding experimental systems using statistical mechanics \cite{binder}. This problem and other issues discussed above, including the increase of $T_g$ with field frequency, remain unresolved at present \cite{binder1}.

Notably, profound non-equilibrium dynamic effects as well as thermodynamic anomalies are observed in both spin and structural glass transitions. Interestingly, the current view of the two phenomena is markedly different. The predominant opinion is that the structural glass transition is not a phase transition, and is a purely dynamic effect despite the heat capacity jump at $T_g$ \cite{dyre,tarjus1,angell,angell1,phil,langer,ngai}. The current view also holds that the spin glass transition is a phase transition, the unconventionality of which comes from the dynamic effects \cite{mydosh,s2,binder,binder1}. For example, Mydosh asks: ``How can we get rid of the dynamical processes to determine the basic properties of the underlying phase transition? At present an answer has not yet appeared'' \cite{mydosh}.

In this paper, we propose that instead of attempting to get rid of universal and profound dynamic effects in order to discuss the underlying phase transition, a more fruitful approach is to consider that the dynamic effects are at the heart of the spin glass transition. In this approach, we explore whether the main experimental results can be understood in an entirely dynamic picture. Contrary to the assumption currently held \cite{s2}, we propose that paramagnetic and spin glass states are qualitatively the same. Consequently, we do not discuss a phase transition of some sort, and therefore do not face complications and controversies surrounding the nature of the spin glass phase and a phase transition \cite{mydosh,s2,s3,binder}. We propose that the central experimental result, the susceptibility cusp, necessarily originates when a system stops relaxing at the experimental time scale. In this picture, the observed cusp marks the {\it crossover} that separates two time regimes, high-temperature relaxational regime $t>\tau$ and low-temperature spin wave regime $t<\tau$, where $t$ is the observation time and $\tau$ is system relaxation time. Time-dependent effects, inconsistent with the phase transition approach, and the logarithmic increase of $T_g$ with field frequency in particular, originate as the immediate consequence of the proposed picture. Our proposal thus represents a new outlook at the problem in which the experimental data can be interpreted in a way that is simpler and more physically transparent, an attractive possibility in view of persisting difficulties and controversies in the field.

In our discussion, we explore close parallels between the spin glass transition and the structural glass transition. Importantly, we do so opposite to the historical trend when the insights from the existing spin glass theories were used to understand the structural glass transition (see, e.g., Refs. \cite{binder1,wol,tarjus,bouch,moore,bert}). These previous approaches invoked the concepts from the spin glass transition theories based on phase transitions and subsequently applied them to the structural glass transition. Consequently, the structural glass transition was related to a phase transition of some sort. On the other hand, we start from a recent picture which explains the structural glass transition as a purely dynamic phenomenon, without any reference to a distinct solid glass phase or a phase transition \cite{ours}. We subsequently propose that the spin glass transition can equally be understood as a purely dynamic phenomenon. The next section outlines the recent dynamic approach to structural glass transition, followed by our proposal that the same general idea can be used to understand the spin glass transition.

\section{Structural glass transition}

\subsection{Dynamic origin of the heat capacity jump at $T_g$}

In the field of spin glass transition, the most commonly discussed effect is the cusp of susceptibility at $T_g$. In the area of structural glass transition \cite{dyre,tarjus1,angell,angell1,phil,langer,ngai}, the response of a system is measured to the heat flow, and is quantified by heat capacity. If a liquid avoids crystallization on cooling, the heat capacity changes with a jump at the glass transition temperature $T_g$. For various liquids, $\frac{C_p^l}{C_p^g}=1.1-1.8$ \cite{angell}, where $C_p^l$ and $C_p^g$ are constant-pressure liquid and glass heat capacities, respectively. The jump of $C_p$ provides one definition of $T_g$ at which the glass is said to form. Because above and below $T_g$ the system structure is equally disordered, the jump of $C_p$ immediately presents a problem that is at the heart of glass transition \cite{dyre,angell}: how can the jump be understood if there is no distinct second phase?

This problem remains unresolved, resulting in several important and interesting controversies \cite{dyre,tarjus1,angell,angell1,phil,langer,ngai}. One set of theories rationalize the jump of heat capacity by relating it to the existence of a phase transition and using the concepts from the phase transitions theory. This approach has been convincingly criticized for a number of reasons (see, e.g., Ref. \cite{dyre1}). Perhaps the most important reason is that it has not been possible to identify a distinct low-temperature phase, the glass phase. To get round this problem, several theories have put forward the proposals about the non-conventional mechanisms of the phase transition and non-trivial descriptions of the second phase, while retaining the idea of a phase transition in general \cite{dyre}. Another set of glass transition theories maintain that glass transition phenomena at $T_g$ have purely dynamic origin, corresponding to the freezing of atomic jumps in a liquid at the experimental time scale \cite{dyre}. The assumed absence of a phase transition and thermodynamic anomalies is supported by the wide experimental observation that the liquid and the glass structures at $T_g$ are nearly identical \cite{dyre}. However, the challenge for these theories is to explain the origin of the heat capacity jump at $T_g$ as well as its large magnitude, which can be of the order of $k_{\rm B}$ per atom.

We have recently proposed \cite{ours} how to explain the jump of heat capacity in a purely dynamic picture, without asserting the existence of a distinct phase and a phase transition of some sort and, therefore, showed how to reconcile the above contradiction. We recall that glass transition temperature $T_g$ has two experimental definitions which give similar values of $T_g$. In the calorimetry experiments, $T_g$ is the temperature at which the jump of $C_p$ is measured. In the experiments that measure liquid relaxation time $\tau$ (e.g. dielectric relaxation experiments), $T_g$ is the temperature at which $\tau$ exceeds the experimental time $t$ of typically $10^2-10^3$ s \cite{dyre}. We proposed that when $\tau$ exceeds $t$, the jump of heat capacity at $T_g$ follows as a necessary consequence due to the change of liquid elastic, vibrational and thermal properties including the bulk modulus and thermal expansion. Therefore, the jump of $C_p$ at $T_g$ is the result of the dynamic crossover rather than a phase transition. Consequently, there is no need to invoke a second glass phase, a phase transition of some sort and face associated problems.

We first discuss why and how liquid thermal expansion coefficient $\alpha$ and bulk modulus $B$ change at $T_g$. Unlike in a solid, atoms in a liquid are not fixed, but rearrange in space. This gives the liquid its ability to flow. Each flow event is a jump of an atom from its surrounding atomic ``cage'', accompanied by large-scale rearrangement of the cage atoms. We call this process a local relaxation event (LRE). A LRE lasts on the order of elementary (Debye) vibration period $\tau_0=0.1$ ps. Frenkel introduced liquid relaxation time $\tau$ as the time between LREs in a liquid at one point in space \cite{frenkel}. Smaller and larger $\tau$ correspond to smaller and larger liquid viscosities, respectively. Basing on this property only, Frenkel concluded that at short times $t<\tau$, liquid response is the same as that of a solid, i.e. is purely elastic. On the other hand, for $t>\tau$, viscous liquid flow takes place, during which each LRE gives the additional, viscous, displacement. Hence, for $t>\tau$, liquid response to external perturbation (e.g. pressure) consists of both elastic and viscous response \cite{frenkel}. This picture provided the microscopic basis for the earlier phenomenological model of Maxwell \cite{maxw}, who proposed to separate elastic and viscous response in his viscoelastic approach to liquids.

Lets consider that a liquid is subject to pressure $P$. Pressure induces a certain number of LREs, which bring the liquid to equilibrium at new external conditions ($P$, $T$) after time $\tau$. As follows from the Maxwell-Frenkel approach, the change of liquid volume, $v$, is $v=v_{el}+v_{r}$, where $v_{el}$ and $v_r$ correspond to solid-like elastic deformation and viscous relaxation process due to LREs, respectively. Lets define $T_g$ as the temperature at which $\tau$ exceeds the observation time $t$. This implies that LREs do not operate at $T_g$ during the time of observation. Therefore, $v$ at $T_g$ is given by purely elastic displacement as in elastic solid. Then, $P=B_l\frac{v_{el}+v_r}{V_l^0}$ and $P=B_g\frac{v_g}{V_g^0}$ above and below $T_g$, respectively, where $B_l$ and $B_g$ are bulk moduli of the liquid and the glass, $V_l^0$ and $V_g^0$ are initial volumes of the liquid and the glass, and $v_g$ is the elastic deformation of the glass. Let $\Delta T$ be a narrow temperature interval that separates the liquid from the glass so that $\tau$ in the liquid, $\tau_l$, is $\tau_l=\tau(T_g+\Delta T)$ and $\frac{\Delta T}{T_g}\ll 1$. Then, $V_l^0\approx V_g^0$ and $v_{el}\approx v_g$. Combining the expressions for $B_l$ and $B_g$, we find:

\begin{equation}
B_l=\frac{B_g}{\epsilon_1+1}
\label{2}
\end{equation}

\noindent where $\epsilon_1=\frac{v_{r}}{v_{el}}$ is the ratio of relaxational and elastic response to pressure.

The relationship between $\alpha_l$ and $\alpha_g$ can be calculated in a similar way, giving \cite{ours}:

\begin{equation}
\alpha_l=(\epsilon_2+1)\alpha_g
\label{3}
\end{equation}

\noindent where $\epsilon_2=\frac{v_r}{v_{el}}$ is the ratio of relaxational and elastic response to temperature.

Eqs. (\ref{2},\ref{3}) describe the relationships between $B$ and $\alpha$ in the liquid and the glass due to the presence of LREs in the liquid above $T_g$ and their absence in the glass at and below $T_g$, as long as $T_g$ is the temperature at which $t<\tau$. We note that $B_g$ and $\alpha_g$ can be called unrelaxed, or non-equilibrium values, of bulk modulus and thermal expansion, respectively, because relaxational response $v_r$ in the glass decays during time $\tau$.

We now relate the jump of constant-pressure heat capacity, $C_p$, to the corresponding changes of $\alpha$ and $B$ at $T_g$. We write $C_p=VT\alpha^2B+C_v$, where $C_v$ is the constant-volume heat capacity. We note that $C_v$ around $T_g$, or at any temperature where $\tau\gg\tau_0$, is essentially due to the vibrational motion because the relative number of diffusing atoms is proportional to $\frac{\tau_0}{\tau}$, and is therefore negligibly small \cite{ours}. Using the Gr\"{u}neisen approximation, we have shown that $C_v=3N(1+\alpha T)$ \cite{ours}.

As discussed above, $\alpha$ and $B$ are different above $T_g$ when $t>\tau$ and below $T_g$ when $t<\tau$ (see Eqs. (\ref{2}-\ref{3})). Therefore, the two different time regimes give different values of $C_p$:

\begin{equation}
t>\tau:~C_p^l=VT\alpha_l^2B_l+3N(1+\alpha_l T)
\label{31}
\end{equation}

\begin{equation}
t<\tau:~C_p^g=VT\alpha_g^2B_g+3N(1+\alpha_g T)
\label{32}
\end{equation}

\noindent where $C_p^l$ and $C_p^g$ refer to liquid and glass, respectively.

Eqs. (\ref{31}-\ref{32}) relate the change of heat capacity to the changes of $\alpha$ and $B$ due to the presence of relaxational response in the liquid and its absence in the glass. Using the experimental values of $\alpha_g$, $\alpha_l$, $B_g$ and $B_l$, we have shown that the calculated values of the liquid and glass $\frac{C_p^l}{C_p^g}$ according to Eqs. (\ref{31},\ref{32}) are in reasonable agreement with the experimental $\frac{C_p^l}{C_p^g}$ \cite{ours}.

Importantly, the jump of heat capacity at $T_g$ in our theory takes place within the same single thermodynamic liquid phase, but below and above $T_g$ the liquid is characterized by different values of $\alpha$ and $B$ due to the freezing of LREs at $T_g$ where the liquid falls out of equilibrium. Therefore, our theory is purely dynamic. In contrast to previous theories of glass transition \cite{dyre}, we do not discuss transitions between distinct thermodynamic phases, even though it may be tempting to invoke phase transitions, conventional or unconventional, in order to explain the origin of heat capacity jump. Instead, the jump of heat capacity in this picture is due to the {\it crossover} between two different regimes, relaxational regime above $T_g$ and elastic regime below $T_g$. As we have shown earlier, the crossover may be sharp enough if small variations of temperature around $T_g$ give large changes of $\tau$, leading to the freezing of LREs in a narrow temperature range and giving the appearance of a phase transition \cite{prb}.

As discussed in Ref. \cite{ours} in detail, our theory predicts no difference of $C_p$ between the liquid and the glass at long times when both systems are in equilibrium. In this case, $B_l=B_g$ and $\alpha_l=\alpha_g$, giving $C_p^l=C_p^g$. We note that reaching such an equilibrium state for common glasses can take astronomical times and longer. For example, lets consider SiO$_2$ glass at room temperature $T_r$=300 K. The activation energy barrier $U$ can be assumed constant, because SiO$_2$ is a ``strong'' liquid \cite{angell}. Then, combining $\tau(T_g)=\tau_0\exp(U/T_g)$ and $\tau(T_r)=\tau_0\exp(U/T_r)$, we write $\tau(T_r)=\tau_0\left(\frac{\tau(T_g)}{\tau_0}\right)^{\frac{T_g}{T_r}}$. Taking $\tau_0=$0.1 ps, $T_g\approx 1500$ K and $\tau(T_g)=10^3$ s, we find $\tau(T_r)=10^{67}$ s, approximately the fourth power of the age of the Universe. Although solid for any practical purpose, SiO$_2$ at room temperature is an equilibrium liquid during times $t>\tau(T_r)$. Consequently, it shows no jump of heat capacity on cooling from high temperature if $t>\tau(T_r)$.

\subsection{Time-dependent effects}

The dynamic origin of the jump of $C_p$ at $T_g$ in our theory explains a well-known effect that $T_g$ logarithmically increases with the quench rate $q=\frac{\Delta T}{t}$, where $\Delta T$ is the temperature interval above $T_g$ (see, e.g., Refs. \cite{loga1,loga2}). According to our theory, the jump of heat capacity at $T_g$ takes place when the observation time $t$ crosses relaxation time $\tau$. This implies that $\tau$ at which the jump of heat capacity takes place is $\tau(T_g)=\frac{\Delta T}{q}$. Combining this with $\tau(T_g)=\tau_0\exp\left(\frac{U}{T_g}\right)$ (here $U$ is nearly constant because $\tau$ is approximately Arrhenius around $T_g$ \cite{jpcm}) gives

\begin{equation}
T_g=\frac{U}{\ln\frac{\Delta T}{\tau_0}-\ln q}
\label{log}
\end{equation}

According to Eq. (\ref{log}), $T_g$ increases with the logarithm of $q$, as widely observed in the experiments. Furthermore, the dependence of $T_g$ on $q$ is consistent with the experimental results \cite{loga1,loga2}). This is the immediate and important result of our theory.

We note that Eq. (\ref{log}) predicts no divergence of $T_g$ because the maximal physically possible quench rate is set by the minimal elementary time $\tau_0$ (Debye vibration period) so that $\frac{\Delta T}{\tau_0}$ is always larger than $q$ in Eq. (\ref{log}).

\section{Spin glass transition}

Our main proposal is that, similar to the jump of $B$, $\alpha$ and $C_p$ in the structural glass transition, the susceptibility cusp in the spin glass transition is not related to the existence of a distinct spin glass phase and a phase transition of any sort. Instead, we propose that the cusp is the result of the {\it crossover} between the two regimes, the high-temperature relaxational regime and low-temperature spin wave elastic-like regime,  when observation time $t$ exceeds system relaxation time $\tau$. In this picture, the crossover is a purely dynamic effect that takes place within the same single phase.

\subsection{Susceptibility above and below $T_g$}

Similar to liquids, we define $\tau$ as the time between two consecutive spin jumps (large-scale spin rearrangements in classical picture or transitions between states with different spin quantum numbers) that play the role of LREs in a liquid. When $t>\tau$ at high temperature, spin LREs are operative, governing the dynamics of the system. In this regime, the system is in equilibrium that is equivalent to the equilibrium relaxational state in a liquid at $T>T_g$. Consequently, the equilibrium statistical mechanics can be applied to spin LREs. In a two-level system of non-interacting spins, the free energy is \cite{kittel}:

\begin{equation}
F=-NT\ln\left(2\cosh\frac{\mu H}{T}\right)
\label{fcurie}
\end{equation}

\noindent where $\mu$ is the spin magnetic moment, $H$ is the applied field and $k_{\rm B}=1$. The magnetic moment $M=-\frac{{\rm d}F}{{\rm d}H}=N\mu\tanh\frac{\mu H}{T}$, which at high temperature becomes $M=\frac{N\mu^2 H}{T}$, giving Curie law for susceptibility $\chi_{\rm C}=\frac{{\rm d}M}{{\rm d}H}$:

\begin{equation}
\chi_{\rm C}=\frac{N\mu^2}{T}
\label{claw}
\end{equation}

Curie behaviour of $\chi$ is observed in spin glass systems at high temperature \cite{mydosh}.

On temperature decrease, a spin-glass system can not find an ordered state due to frustration related to disorder of various types \cite{mydosh}. If $U$ is the activation barrier for a spin rearrangement, $\tau=\tau_0\exp\left(\frac{U}{T}\right)$, where $\tau_0$ is the elementary vibration period. $U$ can be temperature-dependent, as discussed in the next section. Hence, $\tau$ continuously increases on lowering the temperature until condition $t>\tau$ is violated at a certain temperature, $T_g$. When $t=\tau$ at $T_g$, spin LREs become frozen at the experimental time scale. At this point, the system falls out of equilibrium and becomes non-ergodic. Consequently, Eqs. (\ref{fcurie}) and (\ref{claw}) do not apply at and below $T_g$. The remaining excitations in the system are spin waves. Associated, in classical representation, with small-amplitude spin displacements \cite{kittel}, spin waves carry thermal energy in an interacting system, ordered or disordered. In spin glasses, spin waves are analogues of elastic waves propagating in glasses or liquids \cite{frenkel}, and can be similarly localized and damped at short wavelengths \cite{s2}. At large wavelengths, spin waves in spin glasses have been predicted theoretically, with the linear dispersion $\omega=ck$, where $c$ is the speed of magnons \cite{swt1}. This has subsequently been confirmed in theoretical studies \cite{swt2,swt3}, simulations \cite{swt5} and experiments \cite{swe1,swe2,swe3,swe4}. We note that spin waves are present above $T_g$ as well \cite{swe4} but this does not alter our main results because spin waves become increasingly damped at high temperature \cite{swt1}.

Unlike magnetic LREs that fall out of equilibrium at $T_g$, the gas of magnons is in thermal equilibrium. Similar to the free energy of phonons \cite{landau}, the free energy of non-interacting magnons is:

\begin{equation}
F=N\epsilon_0+T\sum\limits_i\ln\left(1-\exp\left(-\frac{\hbar\omega_i}{T}\right)\right)
\label{free}
\end{equation}

\noindent where $\omega_i$ are magnon frequencies and $N\epsilon_0$ is the zero-point energy.

Magnetic moment and susceptibility due to spin waves, $\chi_{sw}$, have been calculated for ordered magnetic structures. For example, using the dependence of magnon frequencies in a ferromagnet on external field $H$, $\hbar\omega_k=2JSk^2a^2+g\mu_{\rm B} H$, where $J$ is the exchange parameter, $S$ is spin and $a$ is lattice constant in Eq. (\ref{free}), gives Bloch law: $M\propto -T^{\frac{3}{2}}$ \cite{ahi,tyabl,chak}.

On the other hand, $\chi_{sw}$ has not been calculated for spin glasses. From Eq. (\ref{free}), $\chi_{sw}$ is

\begin{widetext}
\begin{equation}
\chi_{sw}=\chi_0+\frac{\hbar^2}{T}\sum_i\frac{\exp\frac{\hbar\omega_i}{T}}{\left(\exp\frac{\hbar\omega_i}{T}-1\right)^2}\left(\frac{{\rm d}\omega_i}{{\rm d}H}\right)^2-\hbar\sum_i\frac{1}{\exp\frac{\hbar\omega_i}{T}-1}\frac{{\rm d^2}\omega_i}{{\rm d}H^2}
\label{free1}
\end{equation}
\end{widetext}

\noindent where $\chi_0$ is the susceptibility due to zero-point vibrations.

For spin glasses, field dependencies of $\omega_i$ can be taken as $\omega_l=ck$ and $\omega_t=\sqrt{(ck)^2+\left(\frac{\mu H}{2\hbar}\right)^2}\pm\frac{\mu H}{2\hbar}$, where $\omega_l$ and $\omega_t$ are longitudinal and transverse waves, respectively \cite{swt2}. This gives $\left(\frac{{\rm d}\omega_i}{{\rm d}H}\right)^2=\left(\frac{\mu}{2\hbar}\right)^2$ and $\frac{{\rm d^2}\omega_i}{{\rm d}H^2}=\frac{\mu^2}{4\hbar^2\omega_i}$ for small fields ($H\rightarrow 0$) at which $\chi$ is measured in spin glasses \cite{mydosh}. Next, the linear $\omega\propto k$ relationship for small fields implies the quadratic density of states, $g(\omega)$, as for phonons. $g(\omega)=\frac{6N}{\omega_0^3}\omega^2$, where $\omega_0$ is Debye frequency and the normalization coefficient takes into account that $2N$ transverse waves contribute to the sums in Eq. (\ref{free1}) because field derivatives of the longitudinal frequency are zero and hence do not contribute to $\chi_{sw}$. Using the above field derivatives, substituting the sums in Eq. (\ref{free1}) with integrals with $g(\omega)$ and noting that integration can be extended to $\infty$ at low temperature due to the fast convergence of the integrals, we find:

\begin{equation}
\chi_{sw}=\chi_0+\frac{\pi^2}{4}N\mu^2\frac{T^2}{T^3_{\rm D}}
\label{waves}
\end{equation}

\noindent where $T_{\rm D}=\hbar\omega_0$ is Debye temperature of spin waves.

We find, therefore, that $\chi_{sw}$ is quadratic with temperature and approaches constant value $\chi_0$ at $T\rightarrow 0$, consistent with experimental results (see, e.g., Refs \cite{mydosh,v5,v9,male,tholence1}). Interestingly, the same behaviour (quadratic at low temperature and temperature-independent at $T\rightarrow 0$) was found in the EA theory based on the description of the spin glass transition in terms of a phase transition and order parameters \cite{edwards}.

We note that the approximations used above, including the limit of zero $H$ and the dependence of $\omega$ on $k$ and $H$, may affect the low-temperature behavior of $\chi$. However, it is clear that the behavior of $\chi_{sw}$ below $T_g$ is qualitatively different from $\chi_{\rm C}$ above $T_g$ due to a different mechanism of magnetic response.

\subsection{Dynamic crossover at $T_g$}

As long as spin LREs are operative and in equilibrium when $t>\tau$, Eqs. (\ref{fcurie}) and (\ref{claw}) apply. This gives $\chi=\chi_{\rm C}$ in the long-time relaxational regime above $T_g$:

\begin{equation}
t>\tau: ~\chi=\chi_{\rm C}
\label{curie}
\end{equation}

Regime (\ref{curie}) is equivalent to the high-temperature regime in the structural glass transition where LREs in a liquid are in equilibrium when $t>\tau$. On the other hand, when $t<\tau$ in the short-time spin-wave regime, magnetic response is governed by Eqs. (\ref{free}) and (\ref{waves}). This gives $\chi=\chi_{sw}$ at and below $T_g$:

\begin{equation}
t<\tau: ~\chi=\chi_{sw}
\label{sw}
\end{equation}

Therefore, we find that temperature decrease results in the dynamic {\it crossover} between the two regimes, relaxational (\ref{curie}) and spin-wave regime (\ref{sw}). This crossover is analogous to the crossover between relaxational and elastic regime in liquids, see Eqs. (\ref{31}) and (\ref{32}). We propose that the experimental susceptibility cusp is due to this crossover. We make two further observations supporting this proposal.

First, we show that the crossover at $T_g$ can be fairly sharp, giving the appearance of a ``cusp'' characteristic of true phase transitions. We identified spin glass systems for which the temperature dependence of $\tau$ has been measured, and selected several representative systems in a wide range of $T_g$ between about 3 K and 300 K (see Table 1). In Refs. \cite{v6,v2,v5,v1,v9,v7,v3,v4,v8}, $\tau$ has been fitted to the Vogel-Fulcher-Tammann (VFT) law: $\tau=\tau_0\exp\left(\frac{A}{T-T_0}\right)$. Using the dependence $\tau(T)$, we calculate what temperature increase above $T_g$, $\Delta T_g$, is needed in order to reduce $\tau$ by an order of magnitude: $\tau(T_g)=10\tau(T_g+\Delta T_g)$. $\Delta T_g$ gives an estimation of the temperature range in which the predicted crossover between the high-temperature relaxational regime above $T_g$ and low-temperature spin-wave regime at $T_g$ operates. The values of both $T_f$ and $\Delta T_f$ are given in Table 1. We also list parameters $A$, $T_0$ and $T_g$ used in the calculation.

\begin{table}
\caption{$T_g$, $\Delta T_g$ and $\Delta T_\chi$ for several spin-glass systems.} \vspace{0.5cm}
\begin{tabular}{lllllll}
\hline
System                                                & $A$ & $T_0$ & $T_g$ & $\Delta T_g$ & $\frac{\Delta T_g}{T_g}$ & $\Delta T_\chi$ \\
                                                      & (K) & (K)   & (K)   & (K)          & (\%)                     & (K) \\
\hline
Na$_{0.3}$NiO$_2$$\cdot$1.3H$_2$O \cite{v6}           & 72     &  1.9  & 3.5   & 0.1  & 2.5  & 2 \\
CdCr$_{1.9}$In$_{0.1}$S$_4$ \cite{v2}                 & 114    &  6.1  & 10.6  & 0.4  & 4.2  & 2  \\
U$_2$CuSi$_3$ \cite{v5}                               & 57     &  17   & 19    & 0.2  & 0.9  & 3 \\
CdCr$_{1.8}$In$_{0.2}$S$_4$ \cite{v2}                 & 132    &  14.1 & 21.1  & 1.0  & 4.7 & 2  \\
CuMn (4.6 \%) \cite{v1}                               & 59     &  25.5 & 27.5  & 0.2  & 0.6 & 2  \\
LaMn$_{0.7}$Mg$_{0.3}$O$_3$ \cite{v9}                   & 100    & 27    & 30.5& 0.3  & 1.0 & 4 \\
Fe$_{93}$Zr$_7$ \cite{v7}                             & 300    &  100  & 105   & 0.2  & 1.9 & 10 \\
Cu$_{0.2}$Zn$_{0.8}$Fe$_2$O$_4$ \cite{v3}             & 85     &  100  & 110   & 3.7  & 3.4 & 15 \\
Ni$_{0.7}$Zn$_{0.3}$Fe$_{1.7}$Ti$_{0.3}$O$_4$ \cite{v4}& 50     &  218  & 226  & 4.7  & 2.1 & 75 \\
PMN \cite{v8}                                         & 472    & 291.5 & 311.9 & 2.3  & 0.7 & 10 \\

\hline
\end{tabular}
\end{table}

According to Table 1, $\Delta T_g$ varies in a narrow temperature interval relative to $T_g$, from a fraction of K to several K in systems with small and large $T_g$, respectively, giving $\frac{\Delta T_g}{T_g}\approx 2$\% on average. Consequently, the crossover between the two regimes can be sharp and appear cusp-like. Further, lets define $\Delta T_\chi$ as the approximate experimental temperature range around $T_g$ in which the high-temperature Curie-like behaviour of $\chi$ crosses over to the low-temperature spin-glass dependence. $\Delta T_\chi$ is shown in the last column of Table 1, and we observe that $\Delta T_\chi$ is generally much larger than $\Delta T_g$. This means that in the temperature range where the experimental crossover of $\chi$ operates, $\tau$ increases by many orders of magnitude. This increase is more than sufficient to induce the proposed crossover between free-moving spins in the paramagnetic regime and frozen spins in the spin-wave regime.

Therefore, Table 1 illustrates that (a) small temperature variations around $T_g$ are enough to give order-of-magnitude changes of $\tau$ and induce the proposed crossover between regimes ($\ref{curie}$) and ($\ref{sw}$), and (b) the proposed mechanism can explain the experimental behaviour of $\chi$ because the predicted temperature range of the crossover is consistent with the range in which the experimental crossover (or, if appropriate, the cusp) of $\chi$ develops.

We observe at this point that the character of change of $\chi$ around $T_g$ in many spin glass systems warrants the term ``crossover'' rather than the ``cusp'', as witnessed by smooth and gradual experimental behavior around $T_g$ and large values of $\Delta T_\chi$ in particular \cite{v1,v3,v4,v5,v6,v7,v9,tholence1,bhow1,stringer}. Figure 1 illustrates this point.

\begin{figure}
\begin{center}
\rotatebox{-90}{\scalebox{0.65}{\includegraphics{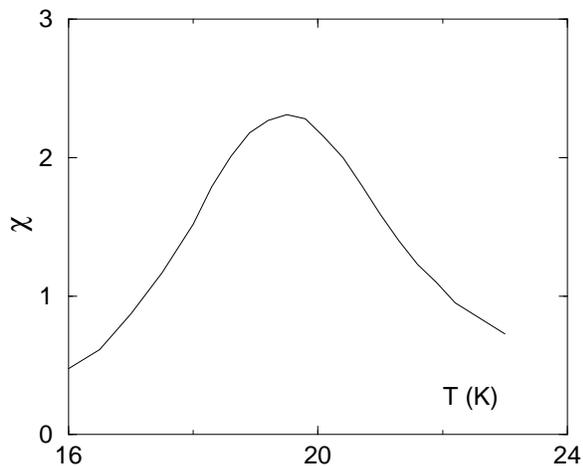}}}
\end{center}
\caption{Experimental spin glass susceptibility \cite{v5} showing the crossover between high- and low-temperature behaviour.}
\label{tg}
\end{figure}

Second, if the observed susceptibility cusp is due to the crossover between two regimes, it is interesting to see that $\chi_{\rm C}$ and $\chi_{sw}$ attain similar values at the crossover temperatures, $T_c$, that are typically observed in the experiment. Order-of-magnitude estimation of $T_c$ can be done by comparing Eq. (\ref{claw}) and Eq. (\ref{waves}) and ignoring system-dependent $\chi_0$ in Eq. (\ref{waves}), a good approximation at higher temperature when $\chi_0$ is small relative to the second temperature-dependent term. This gives $T_c=T_{\rm D}\left(\frac{4}{\pi^2}\right)^{\frac{1}{3}}\approx 0.7 T_{\rm D}$. If typical $T_{\rm D}$ of spin waves are in the range between tens and hundreds of K (e.g. 40--70 K \cite{swe3,swe4}), $T_c$ falls in the range of experimental values of $T_g$ (see Table 1).

In the proposed picture, the susceptibility cusp, or crossover, is related to the freezing of spin LREs at the experimental time scale, similar to the structural glass transition. Consequently, the experimental data can be understood as not a result of a phase transition of some sort between two distinct magnetic phases as discussed previously \cite{mydosh,s2,s3,binder}. Instead, the effects at $T_g$ take place within the same single thermodynamic paramagnetic phase, but the system response crosses over from relaxational above $T_g$ to spin-wave below $T_g$. In this sense, our theory is purely dynamic. In contrast to previous spin glass transition theories \cite{mydosh,s2,s3,binder}, we do not discuss thermodynamic phase transitions and order parameters, even though it may be tempting to invoke phase transitions, conventional or unconventional, in order to explain the cusp of $\chi$. Consequently, our approach involves fewer assumptions and does not require new concepts such as broken ergodicity, ultrametricity, replica symmetry, complexity of the energy landscape and so on in order to explain spin glass transition \cite{mydosh,s2}. Further, in contrast to previous theories, our approach does not face the problems related to the presence of profound relaxation effects in spin glass transition discussed below.

We note that the time-dependent origin of the crossover of $\chi$ implies that the crossover disappears when the spin glass system becomes an equilibrium paramagnet at $t>\tau$. This is analogous to the disappearance of the heat capacity jump in a liquid when $t>\tau$. However, reaching the equilibrium regime at low temperature requires times significantly exceeding the duration of the measurement (see SiO$_2$ example in Section 2).

\subsection{Time-dependent effects}

Spin glass systems universally show profound time-dependent relaxation effects \cite{mydosh,s2,s3,binder,v1,v4,v5,v6,v7,male,bhow1,stringer}. Perhaps the most striking effect is the increase of $T_g$ with the logarithm of field frequency. Inconsistent with phase transition theories, these effects have remained a challenge for spin glass theories based on phase transition and order parameters. To circumvent the problem, these theories added spin dynamics on top of the phase transition framework. This was done by, for example, incorporating relaxation effects into the spin-glass order parameter (see, e.g., Ref. \cite{somp}), introducing spin dynamics according to Glauber or Langevin equations of motion or postulating novel concepts such as hierarchy of relaxation times, energy landscape or ultrametric structure of metastable states, with certain properties attached to each concept \cite{mydosh,s2,s3}.

On the other hand, our dynamic approach does not require any extra steps to rationalize time-dependent effects, and explains time-dependent effects in a simple and straightforward way. This has an added advantage of consistency, in that both the cusp (crossover) of $\chi$ and time-dependent effects are discussed on equal footing.

As proposed, the susceptibility cusp is due to the dynamic crossover between relaxational regime (\ref{curie}), where $t>\tau$, and the spin-wave regime (\ref{sw}), where $t<\tau$. Therefore, the temperature at which the cusp takes place, $T_g$, is defined from the condition $t=\tau$. Lets take $\tau$ in the form of, for example, the VFT law: $\tau=\tau_0\exp\left(\frac{A}{T-T_0}\right)$. Then, noting that observation time $t$ is the inverse of field frequency $\nu$, $t=\tau$ gives

\begin{equation}
T_g=T_0+\frac{A}{\ln\nu_0-\ln\nu}
\label{freq}
\end{equation}

\noindent where $\nu_0=\frac{1}{\tau_0}$ is Debye frequency.

We find, therefore, that the immediate consequence of the proposed theory is the increase of $T_g$ with $\ln\nu$. We further observe that Eq. (\ref{freq}) has precisely the form observed in the experiments (see, e.g., Ref. \cite{v1}). This is the important result of our theory. We note that the increase of $T_g$ with $\ln\nu$ also follows if a simple Arrhenius activation form of $\tau$ is assumed instead of the VFT law, giving $T_0=0$ in Eq. (\ref{freq}).

Eq. (\ref{freq}) is the analogue of Eq. (\ref{log}) which describes the increase of $T_g$ with the quench rate in the structural glass transition. Similarly to the universality of Eq. (\ref{log}), the increase of $T_g$ with $\ln\nu$ according to Eq. (\ref{freq}) is a universal phenomenon, widely observed in spin glasses (see, e. g., Refs. \cite{mydosh,v1,v4,v5,v6,v7,male,bhow1,stringer}).

\subsection{Non-linear susceptibility}

We comment on the behavior of the non-linear susceptibility, $\chi_{nl}=1-\frac{M}{\chi_l H}$, where $\chi_l$ is the linear susceptibility in the limit $H\rightarrow 0$  \cite{mydosh}. Experimentally, $\log\chi_{nl}\propto -\gamma\log\epsilon$, where $\epsilon=\frac{T-T_g}{T_g}$ and $\gamma$ is close to 2 \cite{mydosh,s2,binder}. This is widely taken as the evidence for critical scaling and a phase transition \cite{mydosh,s2,binder}. At the same time, several important problems exist in interpreting this result, including the arbitrariness in choosing $T_g$ and the presence of non-equilibrium state around $T_g$ \cite{s2,binder}. Importantly, the scaling of $\chi_{nl}$ works only away from $T_g$. On the other hand, $\log(\chi_{nl})$ flattens off close to $T_g$ where the scaling fails, inconsistent with a phase transition picture \cite{mydosh,binder}.

The above behavior of $\chi_{nl}$ is qualitatively consistent with the proposed dynamic approach to spin glass transition. Indeed, as discussed above (see Table 1), small temperature increases above $T_g$ result in the rapid decrease of $\tau$. Therefore, the condition for the equilibrium paramagnetic regime, $t>\tau$, starts applying not far above $T_g$. In this regime, $M=N\mu\tanh\frac{\mu H}{T}$, giving $\chi_l=\frac{N\mu^2}{T}$ for small $H$. Using $M$ and $\chi_l$ in $\chi_{nl}$ and expanding $\tanh$ up to the third order in $\frac{\mu H}{T}$, corresponding to the first non-linear term in $\chi_{nl}$, gives

\begin{equation}
\chi_{nl}=\frac{1}{3}\left(\frac{\mu H}{T}\right)^2
\label{nonl}
\end{equation}

Hence, $\log\chi_{nl}\propto -2\log T$. If written as a function of $\epsilon=\frac{T-T_g}{T_g}$, where $T_g$ is an arbitrary temperature, $\log\chi_{nl}=\log\frac{1}{3}\left(\frac{\mu H}{T_g}\right)^2-2\log(\epsilon+1)$. For temperature away from $T_g$ where $\epsilon$ is sufficiently large, $\log\chi_{nl}$ is approximately proportional to $-2\log\epsilon$, as in the experimental data. Closer to $T_g$ where $\epsilon$ is small, $\log\chi_{nl}$ flattens off because $\log(\epsilon+1)$ tends to 0, consistent with the experiments \cite{mydosh,binder}.

We note that Eq. (\ref{nonl}) and the above analysis are valid in the temperature range where $t>\tau$, whereas for smaller temperature and $\epsilon$ they become approximate. For smaller $\epsilon$, $\chi_{nl}$ becomes a non-equilibrium property that describes relaxation of induced magnetization in an experimental time window, and therefore depends on $T$ or $\epsilon$ as is seen experimentally. The detailed discussion of this effect is outside the scope of this paper.

\subsection{Relationship to spin Hamiltonians}

Existing spin glass transition theories typically start with a spin Hamiltonian of the type \cite{mydosh,s2,s3}:

\begin{equation}
H=-\frac{1}{2}\sum\limits_{ij}J_{ij}{\bf S}_i\cdot{\bf S}_j
\label{hami}
\end{equation}

\noindent It is therefore instructive to restate the above idea with a reference to spin Hamiltonians.

Using Hamiltonian (\ref{hami}), spin glass transition theories propose recipes to calculate thermodynamic functions and statistical averages in a spin glass system. Importantly, the EA spin glass order parameter, $q=\langle s_i^{(1)}s_i^{(2)}\rangle$ or more complicated order parameters, are considered to be such statistical averages, and are calculated using the conventional statistical physics formalism \cite{mydosh,s2,s3,binder}. Implicit in this procedure is the assumption that $q$, although originally defined as a dynamic correlator, also describes a distinct equilibrium spin glass phase. On the other hand, we proposed that in order to explain experimental results, this assumption is unnecessary. Consequently, our approach enforces no order parameter and phase transition on the Hamiltonian. We suggested that the spin glass transition is a purely dynamic process where, as in Frenkel's theory of liquids, all correlations decay after time $t>\tau$ \cite{frenkel}. This approach implies $q=0$ at any finite temperature when $t>\tau$.

Notably, our dynamic approach is based on the introduction of $\tau$ from the outset. At the same time, $\tau$ and the subsequent dynamic treatment could in principle be deduced from the first-principles Hamiltonian (\ref{hami}). Indeed, the Hamiltonian specifies all interactions in the system that govern all of its dynamic properties such as spin LREs, their activation barriers and relaxation times as well as spin waves. However, understanding dynamic behaviour of the system based on the description by the first-principles Hamiltonian such as (\ref{hami}) is a very complex task and arguably is not feasible for a real system, particularly for a disordered frustrated system such as the spin glass. In particular, this applies to the determination of relaxation times which are governed by the activation barriers, and which in turn are governed by $J_{ij}$ in a complex way: with modern computers, it is possible to calculate the energy landscape of an interacting system with up to a hundred particles only.

In this sense, the dynamic approach discussed here is a reduction, in that $\tau$ is introduced, but not derived from the general Hamiltonian. Similarly, Frenkel's approach to liquids \cite{frenkel} was a reduction from the first-principles description of a liquid by a general Hamiltonian such as (\ref{hami}), where atomic coordinates $u$ are used instead of spin variables $S$: $H=\frac{1}{2}\sum\limits_{ij}J_{ij}u_i u_j$. This reduction surpassed the above complexity related to the first-principles treatment of the Hamiltonian, and was based on the empirical observation that liquids flow and that $\tau$, or viscosity, is the basic property of the flow.

The introduction of $\tau$ highlights another important difference between the dynamic description of spin-glass transition and previous thermodynamic approaches. The thermodynamic approach ignores the dynamics from the outset: it assumes that the system visits all of its states and therefore considers infinite waiting times. Implicit in this approach is the absence of $\tau$ as the flow property of the system.

\section{Similarities between the structural and spin glass transition: summary}

We have summarized the important features that are common to the structural and spin glass transition in Table 2. Points 1--9 in the Table refer to the similarities between liquids and magnetic systems, including effects at $T_g$. Points 10--12 refer to the similarities of relaxation of both systems above $T_g$, discussed in the next section.

\begin{table*}
\caption{Comparison of the main effects of structural and spin glass transition} \vspace{0.5cm}
\begin{tabular}{lll}
\hline
Property & Structural liquids & Spin glass systems \\
& and glasses &\\
\hline
1. Inability to find an ordered state at low temperature & Yes & Yes\\
2. Relaxation is governed by the dynamics of activated local events & Yes & Yes\\
3. Elastic (spin wave) response at short time $t<\tau$ & Yes & Yes\\
4. Relaxational response at long time $t>\tau$ & Yes & Yes\\
5. The crossover between equilibrium (ergodic) relaxational and non-equilibrium \\
(non-ergodic) elastic regime at $T_g$ when $t=\tau$ & Yes & Yes\\
6. Anomalous behaviour of $C_p$ in liquids and $\chi$ in spin systems at $T_g$ & Yes & Yes\\
7. Logarithmic increase of $T_g$ with cooling rate and field frequency & Yes & Yes\\
8. Distinct glass phase at and below $T_g$ & No & No \\
9. Phase transition at $T_g$ & No & No \\
Phenomena above $T_g$:\\
10. Mediating interactions and propagating waves & Elastic & Magnetic\\
11. At high temperature, relaxation is exponential, non-cooperative and Arrhenius & Yes & Yes\\
12. Crossover to slow non-exponential (stretched-exponential or logarithmic) \\
and non-Arrhenius relaxation at low temperature & Yes & Yes\\
\hline
\end{tabular}
\end{table*}

\section{Relaxation effects above $T_g$}

As in structural glass transition, two main effects that set in above $T_g$ in spin glass systems are the VFT law mentioned above and stretched-exponential relaxation (SER). In liquids, slow relaxation sets in at high temperature $T_c^{\prime}$ where $\tau$ is on the order of picoseconds \cite{cr1,cr2,cr3}. Above $T_c^{\prime}$, relaxation is exponential. It was assumed or postulated that SER is due to some sort of cooperativity of molecular relaxation in a liquid that sets in on temperature decrease, but the physical origin of cooperativity has remained unknown.

We have recently proposed \cite{jpcm} that the physical origin of the VFT law and SER in a liquid is the elastic {\it interaction} between LREs. Large atomic displacements due to the LRE distort the surrounding liquid and induce elastic waves. The wave frequency, $\omega$, is on the order of Debye frequency because the wavelength is on the order of interatomic separations, implying that $\omega>1/\tau$ in the range of $\tau$ relevant for glass transition. As discussed by Frenkel \cite{frenkel}, waves with frequency $\omega>1/\tau$ propagate in a liquid as in a solid. The waves distort cages around other LRE centres in the liquid, and therefore affect their relaxation. Hence, we identified elastic waves as the physical mechanism of mediating interaction, and proposed \cite{jpcm} that this interaction sets the cooperativity of relaxation whose origin was discussed \cite{dyre,tarjus1,angell,angell1,langer,ngai} but not understood from the physical point of view. The key question is the range of this interaction. As discussed in detail \cite{jpcm}, this range is given by $d_{\rm el}$:

\begin{equation}
d_{\rm el}=c\tau
\label{del}
\end{equation}

\noindent where $c$ is the speed of sound.

The non-trivial point is that $d_{\rm el}=c\tau$ {\it increases} with $\tau$. This is directly opposite to the commonly discussed decay of hydrodynamic waves, whose propagation range varies as $1/\tau$. The difference is due to the solid-like regime of wave propagation, $\omega\tau>1$, which is qualitatively different from the hydrodynamic regime, $\omega\tau<1$. We called $d_{\rm el}$ liquid elasticity length because it defines the range over which two LREs interact with each other via induced elastic waves. Importantly, $d_{\rm el}=c\tau$ {\it increases} on lowering the temperature because $\tau$ increases. We proposed that this is the key to the origin of slow relaxation, the VFT law \cite{jpcm} as well as SER \cite{ser}.

As such, the above mechanism is general enough to describe relaxation in any dynamic system where local entities relax and induce mediating waves (elastic, spin or other) that affect relaxation of other entities in the system. Hence, we propose that the same mechanism operates in spin glass systems. Here, a localized spin rearrangement induces a spin wave that affects relaxation of other spins in the system. When $d_{\rm el}<d$, where $d$ is the characteristic distance between the neighbouring spins, spin LREs do not interact, resulting in exponential and Arrhenius relaxation as expected. However, when $d_{\rm el}>d$ on lowering the temperature, interaction between spin LREs sets in. Similar to liquids, we propose that this interaction is responsible for the VFT law and SER seen in spin glass systems \cite{mydosh,s2,v6,v2,v5,v1,v9,v7,v3,v4,v8}.

Eq. (\ref{del}) predicts that, similarly to liquids \cite{jpcm}, two dynamic crossovers in spin glass systems operate at $\tau_1=\frac{d}{c}$ and $\tau_2=\frac{L}{c}$, where $c$ is the speed of spin wave and $L$ is system size. The first crossover is from exponential (non-cooperative) to non-exponential (cooperative) relaxation. The second crossover is from the VFT to a more Arrhenius relaxation when $d_{\rm el}$ reaches system size $L$ \cite{jpcm}. In contrast to structural glass transition, the dynamic crossovers in spin glass systems have not been experimentally studied in detail. Therefore, our prediction can be investigated in future high-temperature experiments in spin glass systems. We note that this picture predicts that $\tau$ and $T_g$ increase with $L$ when $d_{\rm el}\ge L$, provided that the thermalization length of the spin wave is larger than $d_{\rm el}$. Seen in supercooled liquids \cite{si1,si2}, this effect is also seen in spin glass systems \cite{si3,si4,si5,si6,si7}. Consistent with the proposed picture, it also supports the possibility of the second crossover at $\tau_2=\frac{L}{c}$.

\section{summary}

In this paper, we have proposed to take a new outlook at the spin glass transition problem, in view of persisting difficulties and controversies in the field. We proposed to interpret existing and future experimental data in a way that is simpler and more physically transparent. There is certainly more work to be done to clarify how the proposed ideas apply to the wealth of spin glass systems and phenomena.

We have argued that the structural and spin glass transition are similar, but in a different way than previously thought. Instead of asserting that the spin glass transition is related to a phase transition and order parameters and rolling out this approach to the structural glass transition \cite{wol,tarjus,bouch,moore,bert}, we proposed that the structural glass transition can be understood as an entirely dynamic phenomenon, and subsequently suggested that equally can the spin glass transition.

We generally remark that a phase transition between two distinct equilibrium phases results in the change of system properties, a well-studied and understood topic in physics \cite{landau,ma}. However, there are many examples, including in structural and spin glass transition, where the system equally changes many of its important properties at a certain temperature, yet the second phase and the order parameter are not apparent, the nature of the phase transition is not clear and the system shows profound dynamic effects. We have proposed that in this case, the change of system properties can be understood as an entirely dynamic phenomenon with no reference to a phase transition of some sort and order parameters. In this simple picture, system properties change at the dynamic crossover when the system stops relaxing at the experimental time scale when $t=\tau$.

Finally, we have explored closed similarities between relaxation properties of liquids and spin glass systems above $T_g$ and proposed that slow relaxation effects are due to interactions between spin LREs via induced spin waves. This can stimulate future experiments to study high-temperature relaxation effects in spin glass systems.

I am grateful to J. S. McCloy, V. V. Brazhkin, V. Heine, V. Ryzhov, and J. C. Phillips for discussions, and to EPSRC and SEPnet for support.

\end{document}